\documentclass[aps,prl,twocolumn,superscriptaddress]{revtex4-1}
\usepackage{graphicx}
\usepackage[colorlinks,urlcolor=blue,citecolor=blue,linkcolor=blue]{hyperref}
\usepackage{color}

\def\vecsign{\mathchar"017E}
\def\dvecsign{\smash{\stackon[-1.95pt]{\vecsign}{\rotatebox{180}{$\vecsign$}}}}
\def\dvec#1{\def\useanchorwidth{T}\stackon[-4.2pt]{#1}{\,\dvecsign}}
\usepackage{stackengine}
\stackMath

\begin{document}


\title{Accurate Determination of the Dynamical Polarizability of Dysprosium}


\author{C. Ravensbergen}
\affiliation{Institut f{\"u}r Quantenoptik und Quanteninformation (IQOQI), {\"O}sterreichische Akademie der Wissenschaften, Innsbruck, Austria}
\affiliation{Institut f{\"u}r Experimentalphysik, Universit{\"a}t Innsbruck, Austria}

\author{V. Corre}
\affiliation{Institut f{\"u}r Quantenoptik und Quanteninformation (IQOQI), {\"O}sterreichische Akademie der Wissenschaften, Innsbruck, Austria}
\affiliation{Institut f{\"u}r Experimentalphysik, Universit{\"a}t Innsbruck, Austria}

\author{E. Soave}
\affiliation{Institut f{\"u}r Experimentalphysik, Universit{\"a}t Innsbruck, Austria}

\author{M. Kreyer}
\affiliation{Institut f{\"u}r Quantenoptik und Quanteninformation (IQOQI), {\"O}sterreichische Akademie der Wissenschaften, Innsbruck, Austria}
\affiliation{Institut f{\"u}r Experimentalphysik, Universit{\"a}t Innsbruck, Austria}

\author{S. Tzanova}
\affiliation{Institut f{\"u}r Quantenoptik und Quanteninformation (IQOQI), {\"O}sterreichische Akademie der Wissenschaften, Innsbruck, Austria}
\affiliation{Institut f{\"u}r Experimentalphysik, Universit{\"a}t Innsbruck, Austria}

\author{E. Kirilov}
\affiliation{Institut f{\"u}r Experimentalphysik, Universit{\"a}t Innsbruck, Austria}

\author{R. Grimm}
\affiliation{Institut f{\"u}r Quantenoptik und Quanteninformation (IQOQI), {\"O}sterreichische Akademie der Wissenschaften, Innsbruck, Austria}
\affiliation{Institut f{\"u}r Experimentalphysik, Universit{\"a}t Innsbruck, Austria}


\date{\today}

\begin{abstract}
We report a measurement of the dynamical polarizability of dysprosium atoms in their electronic ground state at the optical wavelength of 1064~nm, which is of particular interest for laser trapping experiments. Our method is based on collective oscillations in an optical dipole trap, and reaches unprecedented accuracy and precision by comparison with an alkali atom (potassium) as a reference species. We obtain values of 184.4(2.4)~a.u. and 1.7(6)~a.u. for the scalar and tensor polarizability, respectively. Our experiments have reached a level that permits meaningful tests of current theoretical descriptions and provides valuable information for future experiments utilizing the intriguing properties of heavy lanthanide atoms.
\end{abstract}


\maketitle

\section{}
The dipole polarizability is a quantity of fundamental importance in light-matter interaction, as it characterizes the linear response of a neutral particle to an electric field. The polarizability is related to other important physical quantities, like the van-der-Waals dispersion coefficient, and its knowledge is of great relevance for a deep understanding of many-electron systems, for example in heavy atoms, molecules, and clusters~\cite{bonin1997edp}. The static polarizability characterizes the response to a constant electric field by a single real number. The dynamic polarizability (DP) describes the response to an oscillating field and is represented by a complex frequency-dependent function. Naturally, the DP is much richer and contains much more information on the properties of a particle, in particular on its resonance behavior. While various different methods have been established to measure the static polarizability with high accuracy~\cite{Mitroy2010taa,Gould2005rdi}, measurements of dynamic polarizabilities are notoriously difficult. Accurate laser-spectroscopic methods only provide access to differential polarizabilities, whereas other methods like deflection from a laser beam suffer from the problem of characterizing the interaction region well enough. 

In the realm of ultracold atoms, both the real and imaginary part of the DP play an essential role for controlling the external and internal atomic degrees of freedom. The imaginary part is related to the absorption and scattering of light. The real part gives rise to Stark shifts, which are primarily utilized for constructing optical dipole traps~\cite{Grimm2000odt} in a wide range of different geometries. Zero crossings of the DP, which occur at tune-out wavelengths, can be used to engineer species-selective traps~\cite{LeBlanc2007sso}. Optical lattice clocks operate at a so-called magic wavelength, where the differential DP between the two relevant atomic states vanishes~\cite{Ludlow2015oac}. The DP also enables coherent spin manipulation, which is the basis of many spin-orbit coupling schemes~\cite{galitski2013soc}. 

The optical manipulation of ultracold magnetic lanthanide atoms has attracted considerable interest~\cite{Sukachev2010mot,Lu2011sdb,Aikawa2012bec,Lu2012qdd,aikawa2014oof,Miao2014mot,hemmerling2014bgl,kadau2016otr,Dreon2017oca}.
Their exceptional magnetic properties arise from a partially filled, submerged 4f shell. They feature a very rich atomic spectrum, including narrow optical transitions, and a large orbital angular momentum gives rise to substantial non-scalar contributions to the polarizability. These special properties make magnetic lanthanide atoms excellent candidates to implement advanced light-matter coupling schemes, such as spin-orbit coupling~\cite{burdick2016lls,cui2013sgf}, and to realize novel regimes of quantum matter. The electronic configuration makes advanced calculations of the DP very challenging and interesting~\cite{Dzuba2011dpa,lepers2014aot,Sukachev2016ism,Li2017oto,Li2017aot,golovizin2017mfd}. To benchmark theoretical models, measurements are highly desirable with uncertainties on the percent level. Experimental results have been reported for dysprosium~\cite{Lu2011sdb,Maier2015phd,Schmitt2017phd}, thulium~\cite{Sukachev2016ism,golovizin2017mfd} and erbium~\cite{becher2017apo}, in the latter case also demonstrating the anisotropic nature of the DP. However, all these measurements have been subject to large systematic uncertainties, imposed by the methods at hand.

In this Letter, we report on the accurate determination of the real part of the DP of a magnetic lanthanide atom at a wavelength of particular interest for cooling and trapping experiments. We investigate dysprosium atoms and utilize an idea often applied in precision metrology, performing a measurement relative to a known reference. As a reference species, we use potassium atoms, for which the DP is known on the permille level, and measure the trap frequencies of both species in the same single-beam optical dipole trap (ODT). The frequency ratio is then independent of major experimental systematics and imperfections. In a further set of experiments, we determine the tensor contribution to the DP.

The interaction of atoms with the electric field $\vec{E}$ of laser light is described by the Hamiltonian $H = -\frac{1}{2}\vec{E}^{\dagger}\dvec{\alpha}\vec{E}$, where $\dvec{\alpha}$ is the dynamical polarizability tensor operator~\cite{Deutsch2010qca}. The energy shift for a given quantum state corresponds to the optical trapping potential and is
\begin{equation}
  U(\textbf{r},\omega_L) = -\frac{2\pi a_0^3}{c}I(\textbf{r}) \tilde{\alpha}(\omega_L),\label{FullPolarizability}
\end{equation}
where $\omega_L$ is the laser frequency, $I(\textbf{r})$ the position-dependent intensity, $a_0$ the Bohr radius, and $c$ the speed of light. Here we define $\tilde{\alpha}(\omega_L)$ as a dimensionless quantity corresponding to the real part of the DP of the quantum state of interest in atomic units ($1\,\text{a.u.} = 4\pi \epsilon_0 a_0^3$, where $\epsilon_0$ is the vacuum permittivity). For a Gaussian laser beam, the central region (trap depth $\hat{U}$) can be approximated by a harmonic potential. The corresponding radial trap frequency
\begin{equation}
  \omega_r = \sqrt{\frac{4\hat{U}}{mw_0^2}} = \sqrt{\frac{16a_0^3}{c} \frac{P}{w_0^4}\frac{\tilde{\alpha}(\omega_L)}{m}}\label{TrapFreq}
\end{equation}
is determined by the laser beam parameters (power $P$ and waist $w_0$) and atomic properties (polarizability $\tilde{\alpha}$ and mass $m$)~\cite{Grimm2000odt}. 

The DP can generally be decomposed into the three irreducible contributions $\tilde{\alpha}_S$, $\tilde{\alpha}_V$, and $\tilde{\alpha}_T$ (scalar, vector, and tensor polarizabilty), with weights depending on the angular momentum quantum numbers and the polarization of the trapping light. In our work, we focus on the elementary case of linearly polarized light and atoms in a stretched state~\footnote{Angular momentum projection on the quantization axis equals plus or minus the total angular momentum ($|m_J| = J$)}, where we can decompose $\tilde{\alpha}$ into
\begin{equation}
  \tilde{\alpha}(\omega_L) = \tilde{\alpha}_S(\omega_L) + \frac{3 \text{cos}^2\theta - 1}{2}\tilde{\alpha}_T(\omega_L);\label{TensorPolalizability} 
\end{equation}
here $\theta$ is the angle between the polarization axis and the quantization axis, the latter being defined by the magnetic field. Note that within a hyperfine manifold $\tilde{\alpha}_S$ and $\tilde{\alpha}_T$ only depend on the wavelength.  

The usual method to measure the dynamical polarizability in an ODT~\cite{Lu2011sdb,Maier2015phd,Sukachev2016ism,golovizin2017mfd,becher2017apo} is to determine the trap frequency $\omega_r$ by observing collective oscillations in a trap with a given power $P$ and a well-defined waist $w_0$, and to use Eq.~(\ref{TrapFreq}). A major complication arises from the strong dependence $\tilde{\alpha} \propto w_0^4$. An accurate determination of $w_0$ at the position of the atoms is crucial, but very difficult to achieve in practice. In addition, any aberrations from an ideal Gaussian beam are not accounted for. Moreover, a real cloud with its finite spatial extent will experience some anharmonicity, which will alter the measured oscillation frequency. The combination of these systematic problems typically limits the accuracy of such DP measurements to a few $10\%$~\cite{becher2017apo}. 

\begin{figure}
\includegraphics[width=0.7\columnwidth]{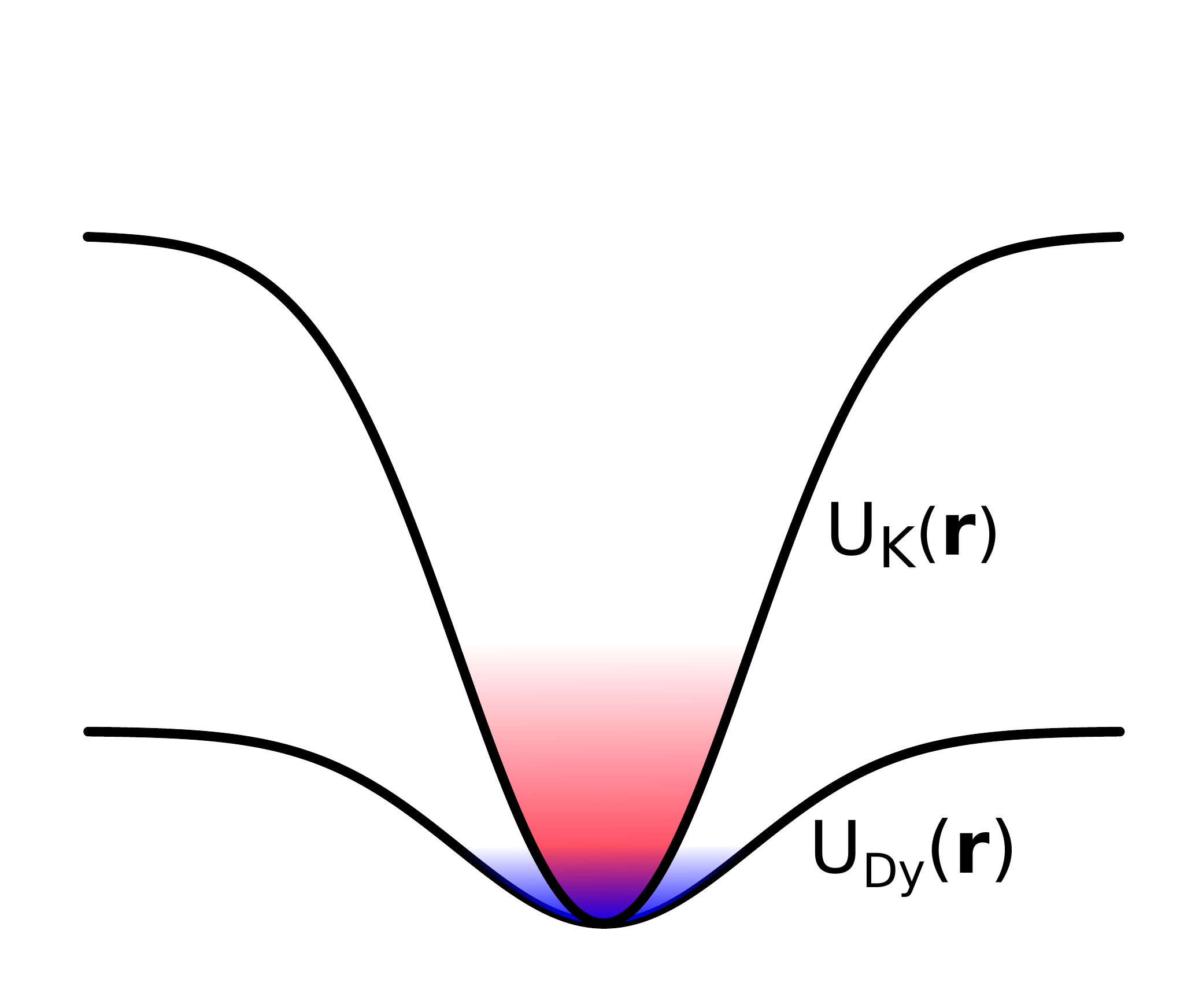}
\caption{Schematic illustration of the species-dependent optical trapping potential $U$ filled with potassium or dysprosium atoms in a beam with a Gaussian profile. Here, in the ideal case, the ratio $T/\hat{U}$ is equal for the two species, the atoms explore exactly the same region in the trap, and thus experience the same anharmonicity and beam aberrations.}\label{Fig1}
\end{figure}

The above limitations can be overcome by referencing the trap frequency of the particle of interest (or state~\cite{Khramov2014uhm}) to a species with a known polarizability~\cite{Neyenhuis2012apo,Danzl2010auh}. Figure~\ref{Fig1} illustrates the situation for two species in the same optical trapping field, where different potential depths result from the different polarizabilities. Within the harmonic trap approximation, the DP of the unknown species, in our case Dy, is then obtained as
\begin{equation}
  \tilde{\alpha}_{\text{Dy}} = \tilde{\alpha}_{\text{K}} \frac{m_{\text{Dy}}}{m_{\text{K}}}\bigg(\frac{\omega_{\text{Dy}}}{\omega_{\text{K}}}\bigg)^2,\label{DPratio}
\end{equation}
where $\tilde{\alpha}_{\text{K}}$ is the polarizability of the reference species (in our case K), and $m_{\text{Dy}}/m_{\text{K}}$ is the known mass ratio. Experimentally, one only has to measure the frequency ratio $\omega_{\text{Dy}}/\omega_{\text{K}}$, which eliminates the need to determine $w_0$ or $P$. This scheme also removes the effects of anharmonicity provided that the ratio of the temperature to the trap depth is the same for both species. In this ideal case, illustrated in Fig.~\ref{Fig1}, the two thermal clouds fill exactly the same region in the trap, and thus experience the same relative effect of anharmonicity. Introducing another species with a different mass may lead to a different gravitational sag and thus to a shift of the frequency ratio. This effect, however, can be suppressed by using a sufficiently deep and tight trap.

In our experiments, we use the isotopes $^{164}$Dy and $^{40}$K, with a mass ratio $m_{\text{Dy}}/m_{\text{K}}=4.102$. For trapping we use the standard near-infrared wavelength of 1064.5~nm. At this wavelength the polarizability of potassium is $\tilde{\alpha}_{\text{K}}=598.7(1.1)$~\cite{Safronova2013mwf,Safronova2013communication,TensorpartAlphaK}. Based on the available theory values for Dy~\cite{Dzuba2011dpa,Li2017oto}, we can estimate $\tilde{\alpha}_{\text{K}}/\tilde{\alpha}_{\text{Dy}}\approx 3.2$ and $\omega_{\text{K}}/\omega_{\text{Dy}}\approx 3.6$.  

We produce a thermal cloud of either $^{164}$Dy or $^{40}$K atoms in a single-beam ODT. For dysprosium, we employ a laser cooling and trapping scheme similar to Refs.~\cite{Maier2014nlm,Dreon2017oca}. After loading the ODT and some evaporative cooling, we typically trap $10^6$ atoms, spin-polarized in Zeeman substate $|J=8,m_J=-8\rangle$, at about 8 $\mu$K. For potassium, after a sub-Doppler cooling stage~\cite{fernandes2012sdl} which also enhances ODT loading, we have $3\times 10^5$ unpolarized~\cite{TensorpartAlphaK} atoms at ${\sim}30\,\mu$K. The trapping laser (Mephisto MOPA 18 NE) operates on a single longitudinal mode, is linearly polarized, and its power is actively stabilized. All measurements reported here are performed with $P=2.5$~W, $w_0 \approx 30\,\mu$m, and a magnetic field strength of 250~mG.  
    
We measure the trap frequencies by exciting a center-of-mass (CoM) oscillation, the so-called sloshing or dipole mode. In a harmonic potential, this mode does not involve a compression of the cloud and the frequency is thus not affected by the interactions within the cloud or by its quantum statistics~\cite{Dalfovo1999tob}. We excite a pure radial sloshing oscillation by displacing the trap position abruptly in the vertical direction using an acousto-optic modulator. The displacement amounts to approximately $2\,\mu$m, which is smaller than the in-trap radial cloud size of about $\sigma_r = 6 \,\mu$m. After a variable hold time we switch off the trap and perform standard time-of-flight (ToF) absorption imaging. The cloud position is extracted from the images by performing a one-dimensional Gaussian fit to a vertical slice taken from the central part of the elongated trap. Both species are imaged using the same optical path and camera. 

\begin{figure}
\includegraphics[width=\columnwidth]{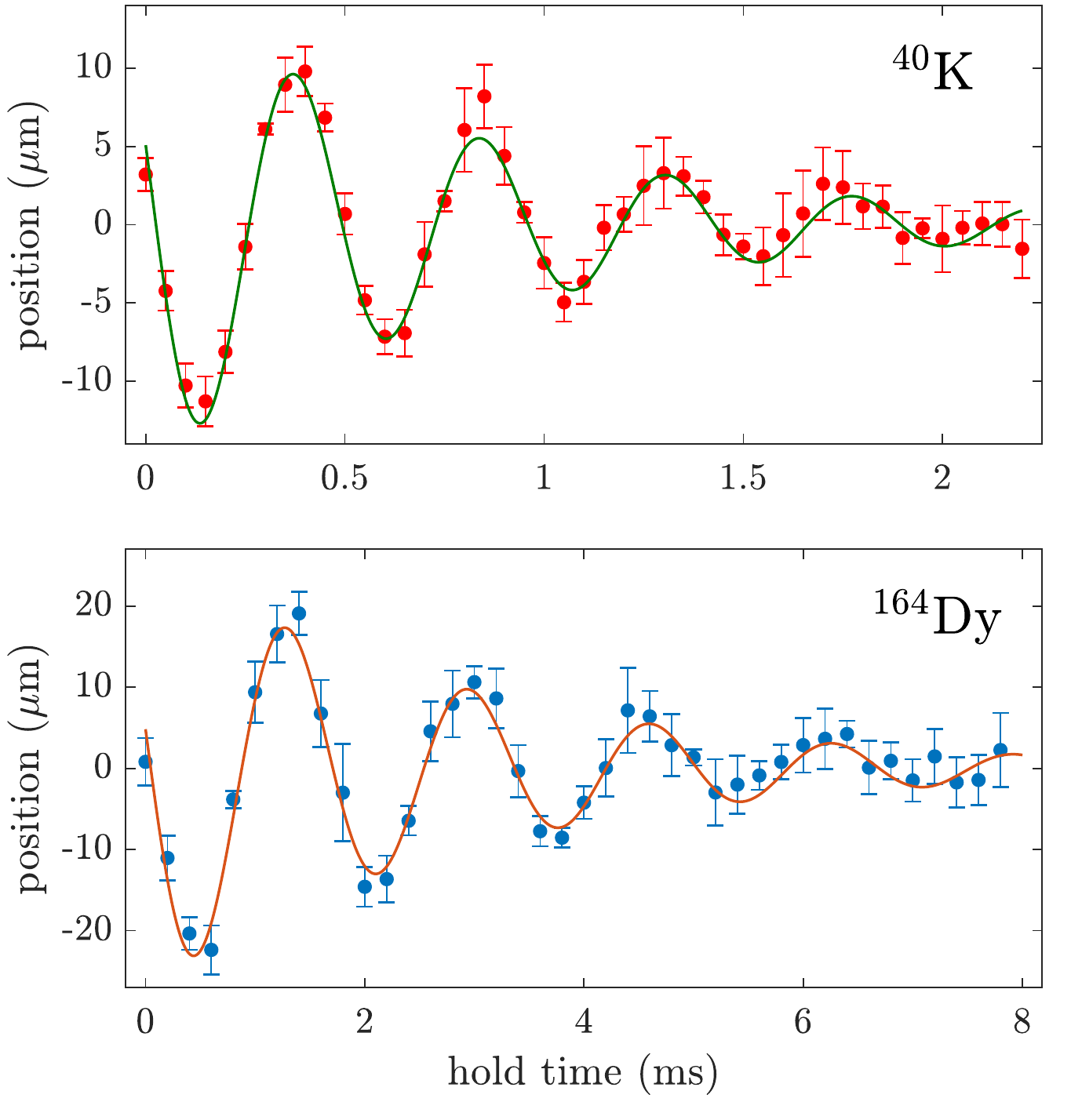}
\caption{Radial sloshing mode oscillation for potassium and dysprosium. The cloud position after ToF is plotted against the hold time in the trap after the excitation. We obtain $\omega_{\text{K}}^{\text{fit}}/2\pi = 2140(10)$ Hz and $\omega_{\text{Dy}}^{\text{fit}}/2\pi = 601(2)$ Hz, $\tau_{\text{K}}=0.8(1)$ ms and $\tau_{\text{Dy}} = 2.9(1)$ ms. The temperatures are $T_{\text{K}}=36(3)\,\mu$K and $T_{\text{Dy}}=8.3(2)\,\mu$K, and the ToF is 0.3 ms for K and 2 ms for Dy. Note that the time scales for K and Dy differ by a factor of 3.6. The error bars show the sample standard deviation of five individual measurements at the same hold time.\label{Fig2}}
\end{figure}

A typical measurement run for both dysprosium and potassium is shown in Fig.~\ref{Fig2}. The magnetic field is chosen to be parallel to the polarization of the trapping light ($\theta=0$), and therefore from Eq.~(\ref{TensorPolalizability}) we get $\tilde{\alpha} = \tilde{\alpha}_S + \tilde{\alpha}_T$. We fit the oscillations with an exponentially damped sine wave to extract the frequency $\omega^{\text{fit}}$ and the damping time $\tau$ of the oscillation. The two species oscillate at different frequencies because of their different mass and polarizability. By relative scaling of the horizontal axes of Fig.~\ref{Fig2} with the expected factor of $3.6$ the oscillations exhibit a nearly identical behavior. This already confirms that the theoretical values of Refs.~\cite{Dzuba2011dpa,Li2017oto} provide a good estimate for the Dy polarizability. The identical damping behavior, with $\omega^{\text{fit}}\tau$ being the same for both species, is consistent with our assumption that the main source of damping is dephasing resulting from the trap anharmonicity~\cite{SupplementalMaterials}.

The measured frequency ratio exhibits a residual anharmonicity effect. After trap loading, plain evaporative cooling reduces the temperature to a certain fraction of the trap depth. This effect is similar, but not exactly equal for both species. We take this into account by a small correction to the dysprosium oscillation frequency. For a given potassium temperature $T_{\text{K}}$ the corresponding dysprosium temperature would be $(\tilde{\alpha}_{\text{Dy}}/\tilde{\alpha}_{\text{K}})T_{\text{K}}$. A deviation from this ideal value can be quantified by $\Delta T_{\text{Dy}}= T_{\text{Dy}}-(\tilde{\alpha}_{\text{Dy}}/\tilde{\alpha}_{\text{K}})T_{\text{K}}$. The anharmonic frequency shift depends on the slope $\beta = \mathrm{d} \omega_{\text{Dy}}/\mathrm{d} T_{\text{Dy}}$, which gives a corrected frequency ratio 
\begin{equation}
\frac{\omega_{\text{K}}}{\omega_{\text{Dy}}} = \frac{\omega_{\text{K}}^{\text{fit}}}{\omega_{\text{Dy}}^{\text{fit}}-\beta \Delta T_{\text{Dy}}}.\label{AnharShift}
\end{equation}
With this correction, Eq.~(\ref{DPratio}) allows to determine $\tilde{\alpha}_{\text{Dy}}/\tilde{\alpha}_{\text{K}}$ in an accurate way.

\begin{figure}
\includegraphics[width=\columnwidth]{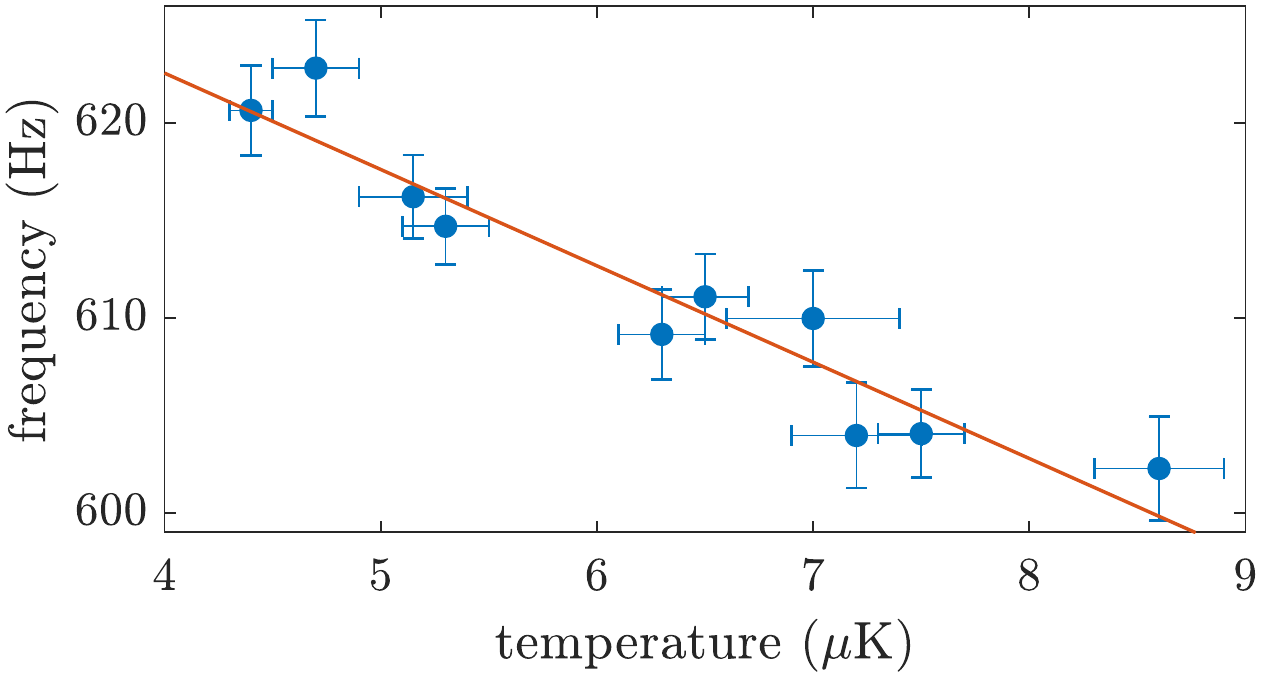}
\caption{Anharmonicity effect on the trap frequency. The Dy CoM oscillation frequency is plotted as a function of the cloud temperature. The weighted linear fit takes both frequency and temperature errors into account, and for the displayed set of measurement yields a slope $\beta/2\pi = -5.1(7)$Hz/$\mu$K.\label{Fig3}}
\end{figure}

In order to determine $\beta$, we vary the temperature of the dysprosium atoms and measure the oscillation frequency. The temperature, determined by standard TOF expansion, is changed by an evaporation ramp down to a variable trap power followed by a re-compression to the standard power and a hold time for thermalization. We observe a frequency decrease with increasing temperature, as is shown in Fig.~\ref{Fig3}.   
From this set of measurements and a second one taken under similar conditions (not shown in Fig.~\ref{Fig3}), we obtain the combined result $\beta/2\pi = -4.5(4)$ Hz/$\mu$K. Note that the anharmonicity shifts the measured Dy frequency, for our typical temperatures and trap depth, by about $5\%$ as compared to the harmonic approximation of Eq.~(\ref{TrapFreq}).

Possible remaining systematics affecting the frequency ratio could include density-dependent interactions, the finite excitation amplitude, and the effect of gravity. We do not observe a density dependence of the oscillation frequency of Dy when varying the atom number over a wide range~\cite{SupplementalMaterials}, confirming that the frequency shift observed in Fig.~\ref{Fig3} can be fully attributed to a change in temperature. The frequency ratio should not be affected by the excitation amplitude, because, for an equal amplitude, both species are affected in the same way. In addition, we varied the excitation amplitude for a single species (Dy) and we did not observe any significant shift for the amplitude used here. The estimated gravitational frequency shift in our trap is $\sim 0.1\%$~\footnote{The relative frequency downshift caused by the gravitational sag can be approximated by: $-2(g/w_0 \omega^2)^2$}, which we neglect in our analysis. Moreover, we noticed that the fitted frequency may slightly depend (on the subpercent level) on the time interval chosen for the analysis. To avoid systematic deviations in the comparison of both species, we choose the time intervals to follow the scaling factor of 3.6. With 0-2.2~ms for K and 0-8~ms for Dy, the intervals then correspond to about twice the respective $1/e$ damping time $\tau_{\text{K}}$ or $\tau_{\text{Dy}}$.\\
\begin{figure}
  \includegraphics[width=\columnwidth]{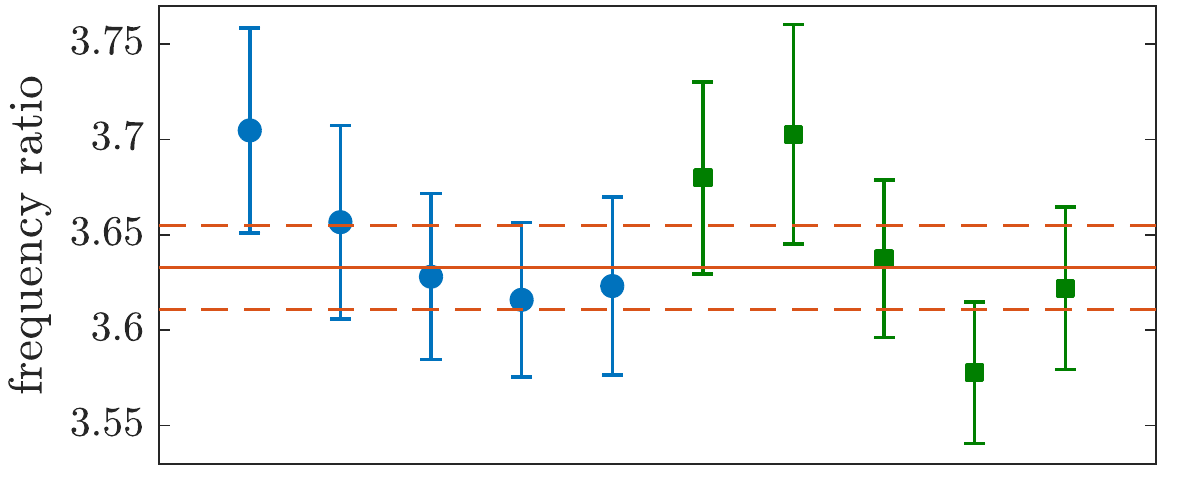}
  \caption{Repeated measurements of the frequency ratio $\omega_{\text{K}}/\omega_{\text{Dy}}$, including small anharmonicity corrections. The two symbols (blue dots and green squares) represent the data sets taken on two different days. The error bars include the fit errors of the frequency measurements and all uncertainties in the anharmonicity correction. Because of the latter, the uncertainties are partially correlated, which we properly take into account in our data analysis when combining the individual results. The solid line markes the final result $\omega_{\text{K}}/\omega_{\text{Dy}} = 3.632(22)$, with the dashed lines indicating the corresponding error range~\cite{SupplementalMaterials}.}\label{Fig4}
\end{figure}
We now turn our attention to an accurate and precise determination of the frequency ratio $\omega_{\text{K}}/\omega_{\text{Dy}}$. We measure the potassium and dysprosium CoM oscillation frequency, in the same trap, in an alternating fashion to eliminate possible slow drifts over time, and repeat this 10 times. The resulting frequency ratios, including small anharmonicity corrections, are shown in Fig.~\ref{Fig4}. The data were taken on two different days, which were one week apart, and the consistency shows the robustness of the presented method. The differential anharmonicity effect from Eq.~(\ref{AnharShift}) yields a small correction of about 1.4$\%$ and 2.2$\%$ for the frequency ratio of the two data sets. The combined result for the frequency ratio is $\omega_{\text{K}}/\omega_{\text{Dy}}=3.632(22)$; for details on the error budget see~\cite{SupplementalMaterials}. 

In a second set of experiments, we measure the frequency ratio $\omega_{\|}/\omega_{\perp}$ for Dy in a magnetic field parallel and perpendicular to the polarization of the laser field. In this way, we can identify the tensor part which is expected to be more than 100 times smaller~\cite{Li2017oto} than the scalar part. Here we perform in total 11 pairs of measurements~\cite{SupplementalMaterials}, alternating the angle $\theta$ between 0 and $\pi/2$. We obtain the combined result $\omega_{\|}/\omega_{\perp} = 1.0070(24)$, which significantly deviates from one and thus reveals a tensor contribution.

From the measured frequency ratios and Eqs.~(\ref{FullPolarizability}-\ref{TensorPolalizability}), it is now straightforward to derive the polarizability ratios $(\tilde{\alpha}_S+\tilde{\alpha}_T)/\tilde{\alpha}_{\text{K}}= 3.217(40)$ and $(\tilde{\alpha}_S+\tilde{\alpha}_T)/(\tilde{\alpha}_S-\tilde{\alpha}_T/2)= 1.014(5)$. Solving for the scalar and tensor part and using the reference value for $\tilde{\alpha}_{\text{K}}$, we finally obtain $\tilde{\alpha}_S=184.4(2.4)$ and $\tilde{\alpha}_T = 1.7(6)$.

Our result for the scalar polarizability lies between the two theoretical values of 180~a.u.~\cite{Dzuba2011dpa} and 193~a.u.~\cite{Li2017oto}, being consistent with both of them within the corresponding er an ror estimates of a few percent~\cite{Dzuba2017communication,Li2017private}. For the small tensorial part, our result is consistent with the theoretical value of 1.34~a.u.~\cite{Li2017oto}.

Already in its present implementation, the experimental uncertainty of our method to determine the DP of a magnetic lanthanide atom is smaller than the uncertainties of theoretical calculations. This, in turn, means that our new result already provides a benchmark and sensitive input for refined theoretical calculations. In extension of our work, much more information on the DP can be obtained by measuring at other optical wavelengths~\cite{becher2017apo}, which is straightforward to be implemented experimentally. Furthermore, experimental uncertainties may be reduced considerably by using the well-defined environment of optical lattices instead of macroscopic trapping schemes. Further advanced DP measurements could provide a wealth of accurate information on the interaction of light with atoms that feature a complex electronic structure, which would go far beyond the present state of the art. 

The presented technique should also be largely applicable to the rapidly expanding field of ultracold molecules~\cite{Carr2009cau,Quemener2012umu}, where diatomic molecules combining alkali and alkaline earth atoms are produced routinely in numerous labs. The increased complexity of the molecular structure, relative to its atomic constituents, renders the precise determination of the dynamic polarizability challenging. Another emerging field aims at direct laser cooling and trapping of more exotic molecules~\cite{Norrgard2016sdm,chae2017odm}, with the benefit of a larger ground state electric dipole moment or applicability to precision measurements. In such systems sympathetic cooling by ultracold alkali atoms~\cite{Hutson2009pfs,Lim2015msc} or even by ultracold hydrogen has been proposed~\cite{Hutson2013uha} as a route to reach quantum degeneracy. In all of the above experiments a spectroscopically well understood species exists either as a constituent forming the molecule or as a coolant, naturally enabling reference measurements of polarizability and other physical quantities.

In our future experiments, we are particularly interested in mass-imbalanced Fermi-Fermi mixtures and possible new superfluid pairing regimes~\cite{Liu2003igs,Iskin2006tsf,Parish2007pfc,Baranov2008spb,Gezerlis2009hlf,Baarsma2010pam,Braun2015zte,wang2017eeo}. For the combination of $^{161}$Dy and $^{40}$K and not far from our present experimental conditions, a ``magic" wavelength is expected to exist where the polarizability ratio for the two species corresponds to the inverse mass ratio. An optical dipole trap operating at this particular wavelength would automatically match the Fermi surfaces of both species after deep evaporative cooling. Based on Refs.~\cite{Dzuba2011dpa} and~\cite{Li2017oto} for Dy and~\cite{Safronova2013mwf} for K, we would expect this wavelength to be at 982nm or 954nm, respectively, and our present measurement suggests it to be in between these two values. The precise location will be subject of further studies.             

%



\begin{acknowledgments}
We thank M. S. Safronova for the data on the K polarizability and for useful discussions. We also thank A. Turlapov, M. Lepers, and O. Dulieu for stimulating discussions. We thank the Innsbruck Er team for kindly sharing their results prior to publication. We acknowledge support by the Austrian Science Fund (FWF) within the Doktoratskolleg ALM (W1259-N27). 
\end{acknowledgments}

%


\begin{thebibliography}{57}
\bibitem [{Ten()}]{TensorpartAlphaK}%
  \BibitemOpen
  \href@noop {} {}\bibinfo {note} {At 1064.5 nm the polarizability of potassium
  is without any significant tensor contribution}\BibitemShut {NoStop}%
\end{thebibliography}
\end{document}